\documentclass[a4paper,traditabstract]{swsc}

\usepackage{graphicx}
\usepackage{txfonts}
\usepackage[displaymath,mathlines]{lineno}
\usepackage[authoryear,round]{natbib}
\usepackage[backref]{hyperref}

\hypersetup{colorlinks=true,citecolor=cyan,urlcolor=cyan,linkcolor=blue}

\begin{document}
	
		
		\title{Modelling solar irradiance from ground-based photometric observations}

		\titlerunning{Modelling solar irradiance from ground-based photometric observations}
		
		\authorrunning{Chatzistergos et al.}
		
		\author{Theodosios~Chatzistergos\inst{1,2} \and
			Ilaria~Ermolli\inst{1} \and
			Fabrizio~Giorgi\inst{1} \and
			Natalie~A.~Krivova\inst{2}  \and
			Cosmin~Constantin~Puiu\inst{1}
		}

		\institute{INAF Osservatorio Astronomico di Roma, Via Frascati 33, 00078 Monte Porzio Catone, Italy 
			\and Max Planck Institute for Solar System Research, Justus-von-Liebig-Weg 3, 37077 G\"{o}ttingen, Germany }
		
		
		\abstract
		{Total solar irradiance has been monitored from space since 1978, i.e. for about four solar cycles. The measurements show a prominent variability in phase with the solar cycle, as well as fluctuations on timescales shorter than a few days. However, the measurements were done by multiple and usually relatively short-lived missions. 
			The different absolute calibrations of the individual instruments and the unaccounted for instrumental trends make estimates of the possible long-term trend in the total solar irradiance highly uncertain.
			Furthermore, both the variability and the uncertainty are strongly wavelength-dependent.
			While the variability in the UV irradiance is clearly in-phase with the solar cycle, the phase of the variability in the visible range has been debated.
			In this paper, we aim at getting an insight into the long-term trend of total solar irradiance since 1996 and the phase of the solar irradiance variations in the visible part of the spectrum.	We use independent ground-based full-disc photometric observations in Ca~II~K and continuum from the Rome and San Fernando observatories to compute the total solar irradiance since 1996.
			We follow the empirical San Fernando approach based on the photometric sum index.
			We find a weak declining trend in the total solar irradiance of
			-7.8$^{+4.9}_{-0.8}\times10^{-3}$ Wm$^{-2}$y$^{-1}$ between the 1996 and 2008 activity minima, while between 2008 and 2019 the reconstructed TSI shows no trend to a marginally decreasing (but statistically insignificant) trend of  -0.1$^{+0.25}_{-0.02}\times10^{-3}$ Wm$^{-2}$y$^{-1}$. The reference TSI series used for the reconstruction does not significantly affect the determined trend. The variation in the blue continuum (409.2 nm) is rather flat, while the variation in the red continuum (607.1 nm) is marginally in anti-phase, although this result is extremely sensitive to the accurate assessment of the quiet Sun level in the images. These results 
			provide further insights into the long-term variation of the total solar irradiance. 
			The amplitude of the variations in the visible is below the uncertainties of the processing, which prevents an assessment of the phase of the variations.
		}

		
		\maketitle
		
		\section{Introduction}

		The solar irradiance (SI) is an important parameter in studies of the solar variability and Earth's climate. 
		The spectrally-resolved SI, called the spectral solar irradiance (SSI), is defined as the solar radiative energy flux per unit area and wavelength as measured at the top of the Earth's atmosphere at a mean distance of one astronomical unit.
		The total solar irradiance (TSI) is the SSI integrated over the whole spectrum. 
		Current knowledge of SI comes from space-based measurements and SI models.

		Measurements collected since 1978 by a series of instruments show that the TSI varies at all discernible timescales from minutes to decades.
		Most prominent are
		a clear change by $\sim$0.1\%~in phase with the solar cycle  \citep{kopp_magnitudes_2016} and fluctuations by up to 0.2--0.3\% on timescales shorter than a few days \citep{domingo_solar_2009}.
		
		On timescales longer than the solar cycle, TSI changes are rather uncertain due to the uncertainties in instrumental calibrations and the limited duration of the individual space experiments. In particular, there are currently four composite records of TSI measurements, presented by \citet{willson_total_1997,dewitte_total_2004,frohlich_solar_2006}; and \cite{dudok_de_wit_methodology_2017}, showing partly conflicting secular trends over the last 4 solar cycle minima. 
		
		The TSI is the integral over the entire solar radiation spectrum that ranges from X-rays to radio waves. The bulk of solar radiant energy is in the visible (Vis, 400--700 nm), followed by the infrared (IR, 700 nm--1 mm) and ultraviolet (UV, 100--400 nm) bands. Measurements of the SSI also started in the 1970s, but are less continuous and more uncertain than those of TSI. Until just less than 20 years ago, they mainly covered the UV band  \citep[see][for a review]{ermolli_recent_2013}. Nevertheless, all present data clearly show that the form and amplitude of SSI variations are strongly wavelength dependent.
		Although the various experiments produced different estimates of the magnitude of UV  changes, all available data agree on that they are in phase with the solar cycle \citep[e.g.][]{floyd_susims_2003,rottman_sorce_2006}. In contrast, SSI measurements at the IR and longer wavelengths show variations that are in anti-phase with the solar cycle.
		In regard to Vis, conflicting results from two experiments make SSI variations at this spectral range still uncertain. In particular, measurements from the Spectral Irradiance Monitor (SIM) on-board the SOlar Radiation and Climate Experiment (SORCE) satellite show Vis changes that are in anti-phase relation with the solar cycle \citep{harder_trends_2009}, while those from the Variability of solar IRadiance and Gravity Oscillations (VIRGO) experiment on-board the SOlar and Heliospheric Observatory (SOHO) display an in-phase variation \citep{wehrli_correlation_2013}.
		However, it has been suggested that the SORCE/SIM data before 2010, which are the ones showing the anti-phase relation, are unreliable due to potential instrumental issues \citep{haberreiter_new_2017,mauceri_revision_2018,mauceri_intercomparing_2020}.

		\begin{figure*}   
			\centering 
					\includegraphics[width=1\textwidth]{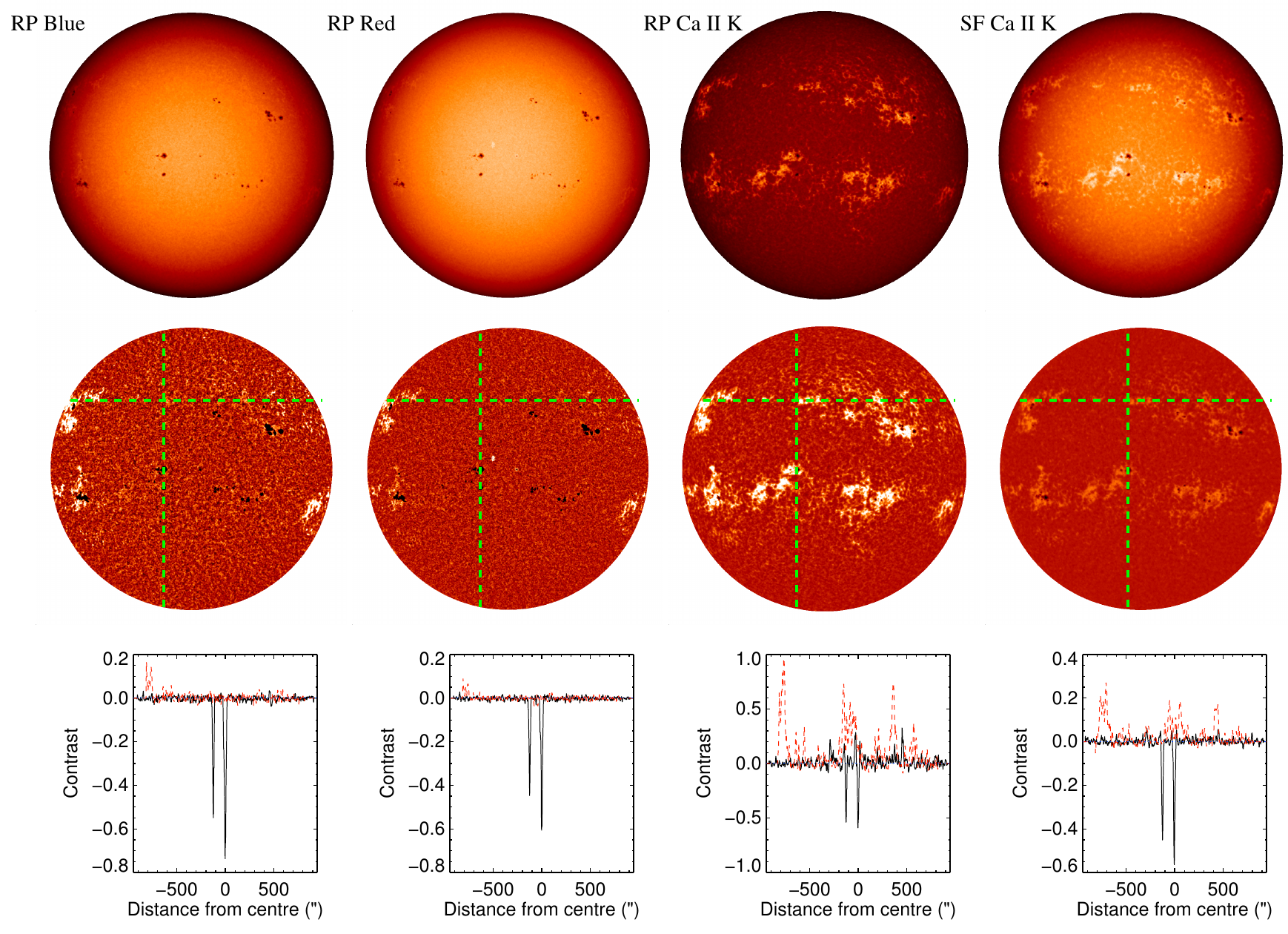}
				\caption{Examples of observations analysed in this study from the Rome/PSPT (RP, columns 1--3) and SFO/CFDT2 (SF, column 4) telescopes. The images were taken on 31 March 2000 in the blue (1st column) and red (2nd column) continuum, and in the Ca~II~K line (3rd and 4th columns). The panels present raw (after the calibration of the CCD) intensity images (top row), limb-darkening-compensated contrast images (middle row),
					as well as horizontal (red) and vertical (black)
					profiles (bottom row) along the cross-sections of the contrast images shown in the middle row, as marked by the dashed green lines. The images have been rotated to display the solar north pole at the top. The raw images are shown to their entire range of values, while the contrast images are shown in the range [-0.05,0.05] and [-0.5,0.5] for the continuum and Ca~II~K images, respectively. In the contrast images, regions outside the ranges listed above are saturated and appear black or white.}
			\label{fig:processedimages_raw} 
		\end{figure*}

		Models that ascribe variations in SI at timescales greater than a day to solar surface magnetism have proved to be successful in reproducing existing measurements of TSI and of SSI at given spectral ranges and timescales \citep[e.g.][]{domingo_solar_2009,shapiro_nature_2017,yeo_reconstruction_2014,yeo_solar_2017}. There are various SI models presented in the literature. They  can be grouped in the so-called  proxy \citep[e.g.][]{hudson_effects_1982,chapman_modeling_2013,tebabal_modeling_2015,georgieva_reconstruction_2015,yeo_empire_2017,lean_estimating_2018,choudhary_variability_2020}, semi-empirical \citep[e.g.][]{fligge_modelling_2000,ermolli_modelling_2001,krivova_reconstruction_2003,crouch_model_2008,shapiro_nlte_2010,fontenla_high-resolution_2011,bolduc_fast_2012,yeo_reconstruction_2014,wu_solar_2018-2}, and physical \citep{yeo_solar_2017} models. 
		The proxy models use linear combinations of indices of solar magnetic features and regression to actual TSI measurements, while semi-empirical and physical models reconstruct the SI variations by employing results from spatially-resolved full-disc observations and from radiative transfer calculations performed on either atmosphere models or outcomes from magnetohydrodynamic simulations of the solar atmosphere, respectively.

		Results from the various SI models show considerable disagreement in the long-term evolution of TSI with estimated changes since the Maunder minimum ranging from 0.7 Wm$^{-2}$ to 6 Wm$^{-2}$ \citep{wang_modeling_2005,krivova_reconstruction_2007,krivova_reconstruction_2010,tapping_solar_2007,steinhilber_total_2009,vieira_evolution_2011,shapiro_new_2011,judge_confronting_2012,judge_sun-like_2020,egorova_revised_2018,wu_solar_2018-2,lockwood_placing_2020}. The various SI models also estimate different SSI variations on timescales longer than a few  solar rotations 
		\citep[see][for a review of this topic]{ermolli_recent_2013}.
		In particular, for the SSI in the Vis range, most models suggest  an in-phase variation  with the solar cycle \cite[e.g.][]{krivova_reconstruction_2006,ball_solar_2011,unruh_solar_2012,yeo_reconstruction_2014,yeo_empire_2017,mauceri_neural_2019}, although there are also models indicating an anti-phase change 
		\citep[e.g.][]{preminger_activity-brightness_2011,fontenla_high-resolution_2011,fontenla_bright_2018}. 
		In the UV, all SI models return variability in phase with the solar cycle.
		
		Long and reliable time series of TSI and SSI are urgently needed to allow accurate quantification of solar contribution to the evolution of Earth's climate.
		To improve our understanding of the solar-cycle and long-term variability of SI,
		here we analyse the TSI and SSI variations in two spectral intervals in the Vis over the last two solar cycles derived from a careful analysis of two independent series of ground-based solar observations. In particular, we exploit the two longest time-series
		of modern full-disc photometric observations of the solar atmosphere, performed at the Rome and San Fernando observatories, and process these data with the  method developed by \cite{chatzistergos_analysis_2018,chatzistergos_analysis_2019,chatzistergos_analysis_2020} to reconstruct the TSI  and SSI variations over the period 1996--2020 by applying the proxy model by \cite{chapman_modeling_2013}.

		The paper is structured as follows.
		In Sect.~\ref{sec:data} we describe the various full-disc solar observations and irradiance series used in our study, as well as the methods employed to process the images and to reconstruct SI variations. In Sect.~\ref{sec:phase} we analyse the solar cycle variability in different spectral bands covered by the full-disc observations and their phase relations.
		We present our results for the long-term variation of TSI over the last three solar activity minima (Sect.~\ref{sec:tsi}). We discuss the results in Sect.~\ref{sec:discussion}, before summarising them and drawing our conclusions in Sect.~ \ref{sec:conclusions}.

		\begin{figure}
			\centering
	\includegraphics[width=1\columnwidth]{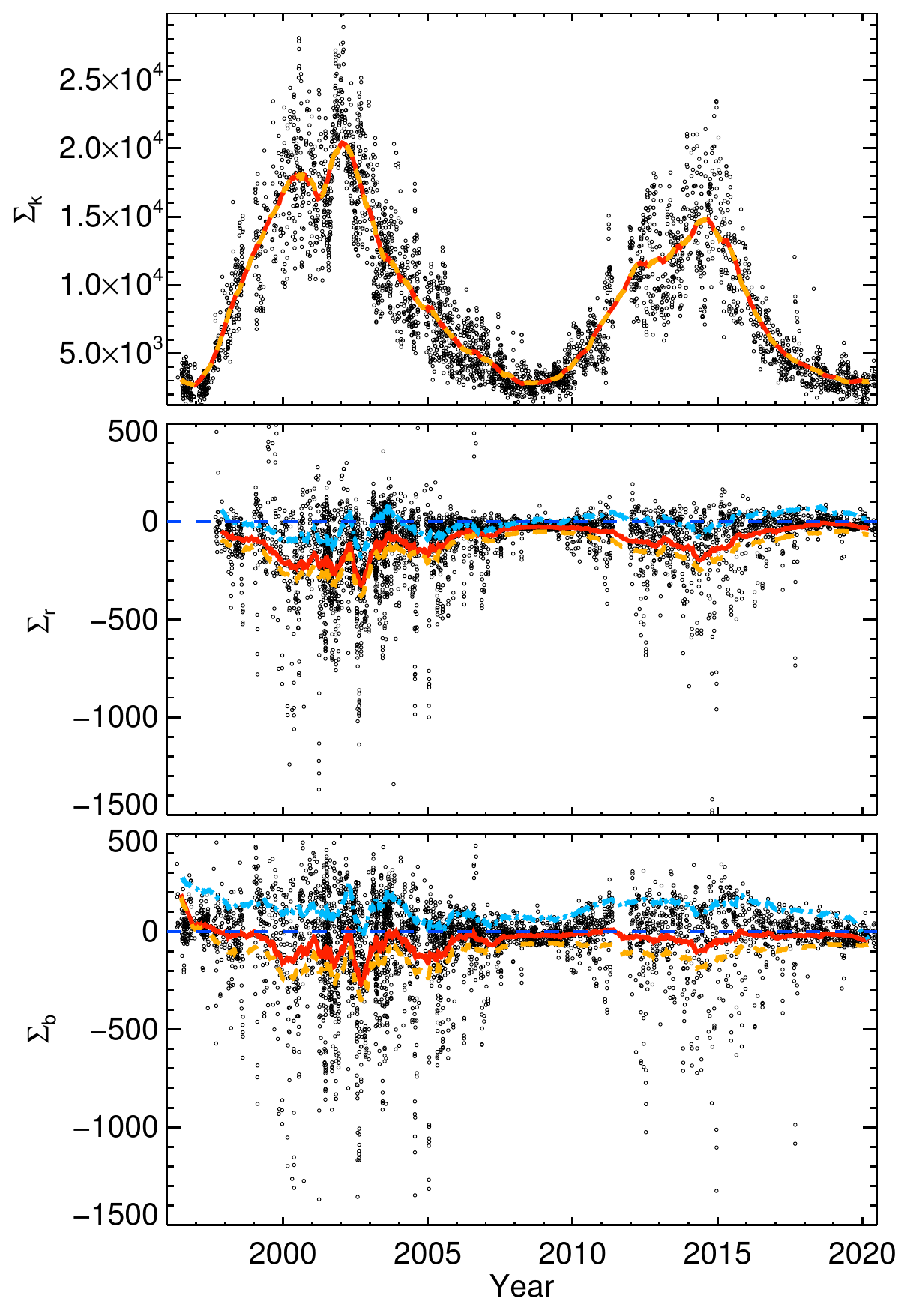}
				\caption{\small Photometric sums from RP Ca~II~K (top), red (middle), and blue (bottom) observations. Daily values are shown in black, while 180-day running averages are shown in red. We note that the photometric sum indices were adjusted, following the method by \cite{nesme-ribes_fractal_1996}, to take into account the bias of the QS level (see Section \ref{sec:phase}). The dashed orange line shows the 180-day running average of the photometric sums prior to this adjustment, while the dash-dotted ciel line (only for the continuum data) was obtained following the approach by \cite{preminger_photometric_2002}. The dashed blue horizontal line denotes the 0-level of photometric sum.}	
			\label{fig:phsumrp1pmod}
		\end{figure}

		\newcounter{tableid}
		\begin{table*}
			\caption{List of datasets analysed in this study.}
			\label{tab:observatories} 
			\centering
			\begin{tabular}{l*{8}{c}}
				\hline\hline
				Observatory & Acronym&	Exposures &Wavelength&SW  &Period		  &  Pixel scale			  	  & Images &Ref.\\
				&   		       & $\#$ of frames&[nm]&[nm]		&  &[$"/$pixel] 			  	  &  &	\\
				\hline
				&   	&  & 393.4&0.25&1996--2020  & 2.0  & 3513  &\addtocounter{tableid}{1}\thetableid\\
				Rome PSPT		     &RP   	&25& 409.2&0.25&1996--2020  & 2.0   & 4367 &\thetableid\\
				&   	&  & 607.1&0.5&1997--2020  & 2.0   & 4135 &\thetableid\\
				\hline
				Rome PSPT				     & RPS  	&1  & 393.4&0.25&2000--2020  & 2.0  & 4230  &\thetableid\\
				&   	&  & 607.1&0.5&2000--2020  & 2.0  & 3766  &\thetableid\\
				\hline
				San Fernando CFDT2	 &SF	& 2& 393.4&0.9&1996--2013  & 2.6 & 3380	  &\addtocounter{tableid}{1}\thetableid\\ 
				\hline
			\end{tabular}
			\tablefoot{Columns are: name of the observatory, abbreviation used in this study, number of summed frames, central wavelength, spectral width, and period of observations, average pixel scale of the images, total number of available images, and the bibliography entry. }
			\tablebib{\addtocounter{tableid}{-\thetableid}
				(\addtocounter{tableid}{1}\thetableid) \citet{ermolli_photometric_2007};
				(\addtocounter{tableid}{1}\thetableid) \citet{chapman_modeling_2013}\textcolor{red}{.}
			}
		\end{table*}

		\section{Data and methods}
		\label{sec:data}
		
		\subsection{Photometric sums and irradiance modelling}
		\label{sec:method}
		
		We follow the approach by \cite{chapman_comparison_2012,chapman_modeling_2013}, who used an empirical model based on the so-called photometric sums derived from  Ca~II~K and continuum images. 
		The assumption is that the irradiance variability is due to the modulation by dark and bright features, such as sunspots and faculae (in white light). Faculae are much better seen in chromospheric  Ca~II~K observations, where they have a higher contrast than in photospheric observations and are called plage. We will refer to both faculae and plage as faculae, independently of the type of their observations.
		Sunspots have a higher contrast in the VIS and IR and therefore images in the red continuum in the 600--700 nm range are used to describe their effect. Faculae are much better represented by  Ca~II~K observations \citep[see e.g.][]{skumanich_sun_1984}.
		
		The photometric sum is defined as $\Sigma=\sum_i(C_iI_i^{\mathrm{CLV}})$. Here $C_i$ is the contrast value of pixel $i$ defined as $C_i=(I_i-I_i^{\mathrm{QS}})/I_i^{\mathrm{QS}}$, with $I_i$ being the intensity value of this pixel and $I_i^{\mathrm{QS}}$ the intensity of the quiet Sun (QS) regions. $I_i^{\mathrm{CLV}}$ is the average QS centre-to-limb variation (CLV) normalised such that at the disc centre it has the value of  unity.
		Thus $C_i$ is dimensionless, and the photometric sum of quiet sun regions is equal to 0.  
		By normalising $I_i^{\mathrm{CLV}}$ to unity we render the photometric sums derived from Ca~II~K and continuum observations analysed here directly comparable.
We note, that the photometric sums as defined here depend on the pixel scale of the images.
		The images and their processing are described in the next section.

		The TSI is calculated as a linear combination of the photometric sums in the red, $\Sigma_r$, {\em or} blue, $\Sigma_b$, continuum, and the one in the  Ca~II~K line, $\Sigma_k$ \citep{chapman_modeling_2013}:
		\begin{equation}
		\label{eq:tsi}
		\mathrm{TSI}(t)=a\left(1+b\Sigma_{r,b}(t)+c\Sigma_k(t)\right).
		\end{equation}
		The	three parameters, $a, b$ and $c$ of the model are  determined through a linear regression of the photometric sums to the actual TSI measurements described in Sect.~\ref{sec:tsirefseries}

		\subsection{Full-disc observations}
		We analysed full-disc solar observations acquired with the Precision Solar Photometric Telescope at the INAF Osservatorio Astronomico di Roma (Rome/PSPT, hereafter) and with the Cartesian Full-Disk Telescope 2 at the San Fernando observatory (SFO/CFDT2, hereafter).
		Table \ref{tab:observatories} lists key characteristics of the analysed full-disc observations.
		In particular, we considered Rome/PSPT observations taken in the Ca~II~K line at 393.37 nm (0.25 nm bandwidth), blue continuum at 409.20 nm (0.25 nm bandwidth; blue hereafter), as well as in the red continuum at 607.1 nm (0.5 nm bandwidth; red hereafter). The SFO/CFDT2 filtergrams were acquired in the Ca~II~K line at 393.37 nm (0.9 nm bandwidth).
		
		Regular monitoring of the solar atmosphere in Ca~II~K and the blue continuum with Rome/PSPT\footnote{Available at \url{https://www.oa-roma.inaf.it/pspt-in-rome/}} started in May 1996. 
		Observations in the red continuum started in September 1997 \citep{ermolli_rise-pspt_1998,ermolli_photometric_2007}. 
		Two significant instrumental changes occurred in August 1997 and September 2001, involving the replacement of the CCD device and the Ca~II~K filter, and the telescope relocation from the Monte Mario to Monte Porzio Catone site, respectively. These changes introduced slight discontinuities in the series of the Rome/PSPT observations, which seem not to significantly affect the high photometric accuracy of the Rome/PSPT data  \citep{ermolli_recent_2011,chatzistergos_analysis_2019}. The observations are stored as 1024$\times$1024 pixel$^2$ images with 2\arcsec/pixel size. 
		They are available as single exposure images (RPS, hereafter), as well as the sum of 25 exposures (RP, hereafter).
		The RPS data have better spatial resolution, while the RP images have an improved photometric accuracy. 
		In this study we considered the RP data obtained from 16 May 1996 to 12 June 2020, and the RPS images acquired from 21 August 2000 to 12 June 2020. Even though RP images are sums of 25 exposures, individual exposures were not stored before the 21st of August of 2000.

		The SFO/CFDT2 observations\footnote{Available at \url{http://www.csun.edu/SanFernandoObservatory/sfosolar.html}} analysed in this study cover the period 1996--2013. Regular solar monitoring started at the San Fernando Observatory with the Cartesian Full Disk Telescope 1 (SFO/CFDT1) in 1986. 
		This telescope produces 512$\times$512 pixel$^2$ images with $\sim$5.1\arcsec/pixel size.
		A second telescope, SFO/CFDT2, was introduced in 1992 \citep{chapman_modeling_2013} providing higher spatial resolution data.
		The images acquired with SFO/CFDT2 have dimensions 1024$\times$1024 pixel$^2$ with $\sim$2.6''/pixel size. 
		The images are the sum of two exposures taken roughly 7.5 minutes apart \citep{chapman_modeling_2013}.
		Here, we restricted our analysis to the SFO/CFDT2 data (SF, hereafter), because of their higher resolution compared to SFO/CFDT1, as well as potential instrumental issues with SFO/CFDT1 images \citep{chatzistergos_analysis_2020}.
		We also restricted our analysis to SFO/CFDT2 data taken since 1996 due to the availability of the Rome/PSPT blue continuum observations.

		\begin{figure*}
			\centering
			\includegraphics[width=1\textwidth]{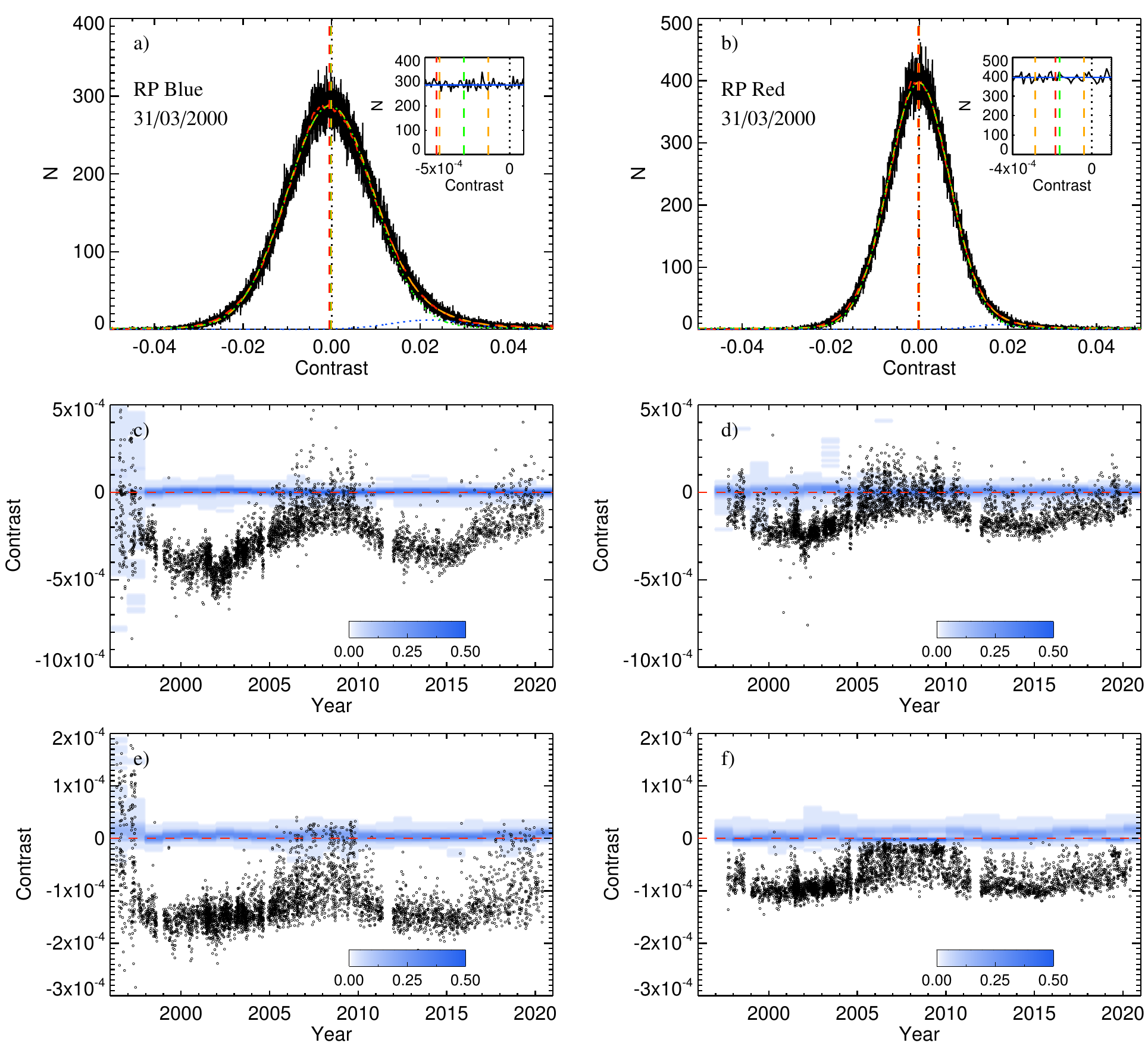}
				\caption{\textit{Panels a and b: }Histogram of pixel values (black) of the RP blue (left) and red (right) contrast images  shown in the middle row of Fig. \ref{fig:processedimages_raw}.
					The fit of the sum of two Gaussians to the histogram of contrast values is shown in orange, while the individual Gaussian functions are shown in green and blue. Also shown in red is the 100-point running mean of the histogram. 
					The insets show enlargements of the central parts to highlight the differences between the various quantities discussed in Sect. \ref{sec:phase}. In particular, the dashed vertical lines show the location of the central Gaussian function for this particular image (red) as well as the mean (green) and the 1$\sigma$ interval (orange) from all RP images, respectively. The black dotted vertical line denotes zero contrast. 
					\textit{Panels c--f: } Temporal variation of the QS level determined with the methods by \citet[][panels c and d]{preminger_photometric_2002} and by \citet[][panels e and f]{nesme-ribes_fractal_1996} for all RP blue (left) and red (right) observations. The dashed red line marks the zero contrast level. Also shown are PDFs of the uncertainty of the QS bias estimations due to the pixel-by-pixel errors of the image processing as evaluated with a Monte-Carlo simulation (see Sect. \ref{sec:phase} for more information). The PDFs are displayed in bins of 1 year and $10^{-5}$ in contrast and they are colour coded as indicated by the colour bars.}
			\label{fig:gaussfit}
		\end{figure*}

		\begin{figure*}
			\centering
			\includegraphics[width=1\textwidth]{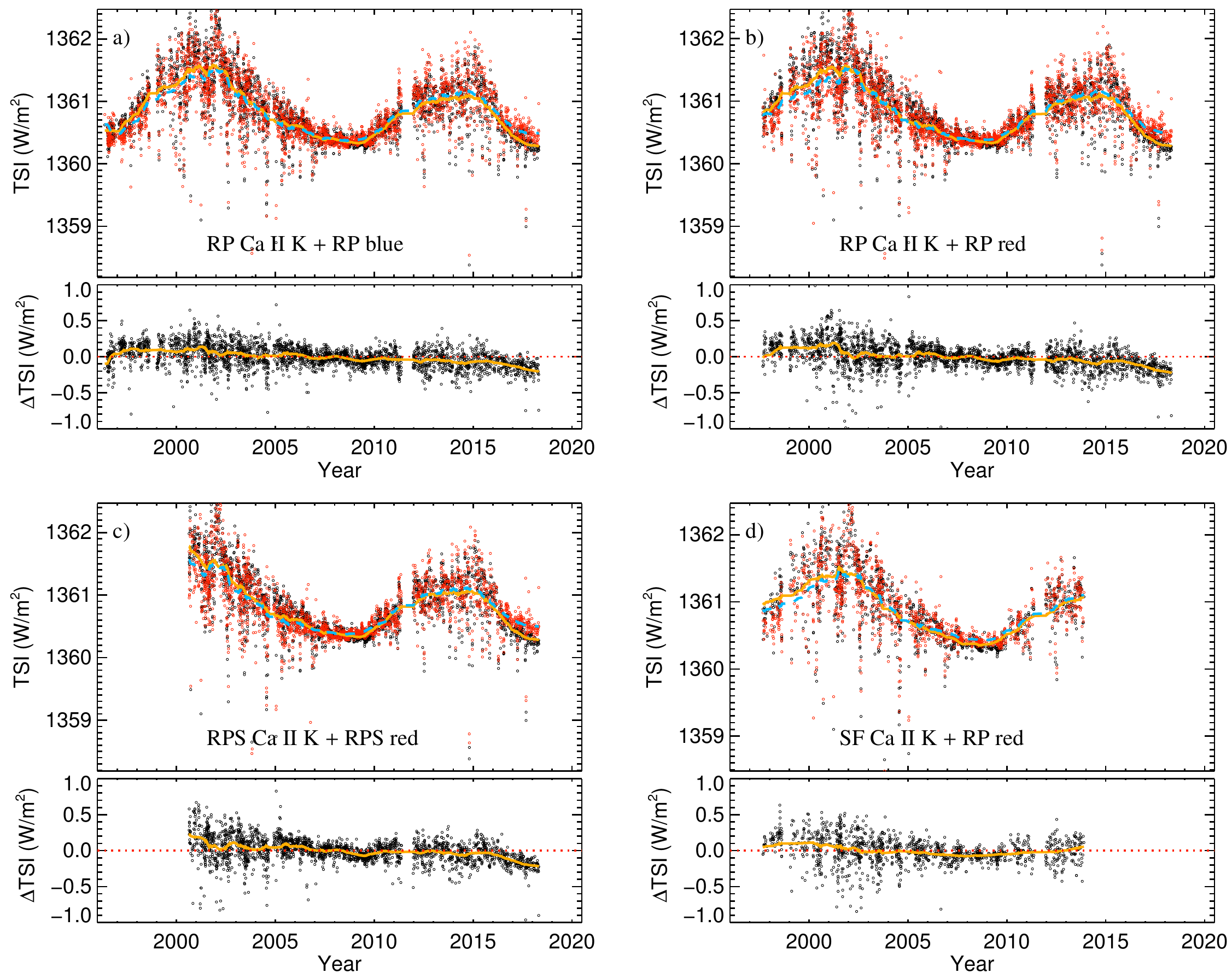}
				\caption{\small 
						Comparison between the reconstructed TSI series (red circles for daily values and dashed ciel line for the 180-day running mean in the top part of each panel) and the PMOD TSI composite series (black circles for daily values and solid orange line for 180-day running mean values). At the bottom of each panel, the corresponding difference between the PMOD TSI and the reconstructed TSI (i.e. PMOD minus our reconstruction) is shown (black circles for daily values and orange line for the 180-day running mean).
						TSI series are reconstructed  using the following data: RP Ca~II~K and RP blue  (a); RP Ca~II~K and RP red (b); RPS Ca~II~K and RPS red (c); and SF Ca~II~K and RP red (d) observations. For all reconstructions, the PMOD TSI composite was used as the reference when deriving the regression coefficients.}
			\label{fig:tsirp1pmod}
		\end{figure*}
		
		All observations were calibrated by the observing teams for the instrumental response.
		We have processed the images with the methods described by \citet[][hereafter referred to as ``our'' method]{chatzistergos_exploiting_2016,chatzistergos_ca_2018,chatzistergos_analysis_2018,chatzistergos_historical_2019,chatzistergos_analysis_2019,chatzistergos_delving_2019,chatzistergos_analysis_2020,chatzistergos_historical_2020}. 
		This allowed us to consistently compensate the RP, RPS, and SF data for the intensity CLV of the QS regions and to account for any artificial intensity patterns affecting the data up to 0.99$r$, where $r$ is the solar disc radius. By this compensation we obtained contrast images. A slightly reduced disc area up to 0.98$r$ was used for the computation of the photometric sums in order to remove the effect of any possible artefacts very close to the limb.
		
		Figure  \ref{fig:processedimages_raw} 
		shows examples of the intensity images analysed in our study, as well as the corresponding contrast images obtained from our processing.
		Also shown (bottom panels) are
		profiles along the horizontal and vertical cross-sections of the contrast images (marked by the dashed green lines in the contrast images).
		In particular, these profiles traverse facular and sunspots regions on the contrast image.
		The contrast of faculae in the RP continuum is clearly lower than in RP Ca~II~K observations.
		In the RP blue,
		the contrast is higher than in the RP red.
		In both RP continuum bands, the contrast in sunspots is considerably higher than in faculae.
		Sunspot contrast is also higher compared to that of faculae in SF Ca~II~K, though the difference is less pronounced than in the RP continuum data.
		In the RP and RPS Ca~II~K data, the facular contrast is higher than the sunspot one. 
		The facular contrast in RP and RPS data is higher than in SF  Ca~II~K data.
		These relations between contrast values in the various data are consistent with the narrower bandwidth used for the RP and RPS observations compared to that employed for the SF data.

		\subsection{TSI series}
		\label{sec:tsirefseries}
		In this study, we considered two original instrumental TSI measurement series, three TSI composites, and three model TSI reconstructions.
		The original instrumental records are the daily TSI measurements  from the SORCE/TIM\footnote{Available at the LISIRD archive at \url{lasp.colorado.edu/lisird/}} \citep[version 18, February 2020;][]{kopp_total_2005-1} and SOHO/VIRGO\footnote{Available at \url{https://www.pmodwrc.ch}} \citep[version 6.5, May 2018;][]{frohlich_-flight_1997} covering the time intervals from 2003 to 2020 and from 1996 to 2018, respectively. The considered TSI composites are the ACRIM\footnote{Available at \url{http://www.acrim.com}} \citep[Active Cavity Radiometer Irradiance Monitor, which is the instrument taken as the reference by][]{willson_total_1997}, PMOD\footnotemark[4] \citep[named after Physikalisch-Meteorologisches Observatorium Davos; version 42.65;][]{frohlich_solar_2006}, and RMIB\footnote{Available at \url{ftp://gerb.oma.be/steven/RMIB_TSI_composite/}} \citep[named after Royal Meteorological Institute of Belgium, in french called IRMB;][]{dewitte_total_2004,dewitte_total_2016}, covering the periods 1978--2013, 1978--2018, and 1984--2020, respectively. 
		These TSI composites are created by daisy-chaining, i.e. using the daily measurements of one instrument as the reference to calibrate the daily records of the second one, which then acts as the reference for the next one, etc. 
		Finally, to compare the results of our study with others, we also 
		considered the daily TSI reconstructions with the SATIRE-S\footnote{Available at \url{http://www2.mps.mpg.de/projects/sun-climate/data.html}} \citep[Spectral And Total Irradiance REconstruction; ][]{yeo_reconstruction_2014},
		EMPIRE\footnotemark[7] \citep[EMPirical Irradiance REconstruction; ][]{yeo_empire_2017},
		and NRLTSI\footnote{Available at \url{https://www.ncei.noaa.gov/data/total-solar-irradiance/}} \citep[Naval Research Laboratory TSI; ][]{lean_estimating_2018} models.

		All these models describe the irradiance variability through the varying contributions of sunspot darkening and facular brightening. 
		SATIRE-S is a semi-empirical model.
		It employs the areas and positions on the solar disc of the various features at a given time as well as their time-invariant brightness contrasts as input.
		The brightness contrasts as a function of wavelength and position on the solar disc are obtained with the radiative transfer simulations from semi-empirical atmospheric models \citep{unruh_spectral_1999}.
		The disc area coverage by the surface features is
		derived from magnetograms and continuum filtergrams acquired from ground- and space-borne telescopes since 1974 \citep[see e.g.][for more details]{yeo_reconstruction_2014}.
		EMPIRE and NRLTSI are empirical models that linearly combine proxies of sunspot darkening and facular brightening
		to match the model outcome to the observed TSI changes. Whereas NRLTSI uses the ordinary least squares regression, EMPIRE uses the orthogonal distance regression accounting for the error in both variables.
		Both models use the Mg II index and sunspot areas as input over the period considered in this study.

		\section{Variation of the photometric sums with the solar cycle}
		\label{sec:phase}
		Figure \ref{fig:phsumrp1pmod} 
		shows the $\Sigma_k$, $\Sigma_b$, and $\Sigma_r$ series obtained from the processing of the  RP Ca~II~K, blue, and red images.
		All of them vary with the solar cycle, although with different amplitudes. In particular, 
		the amplitude of the variation of $\Sigma_b$ and $\Sigma_r$ is considerably lower than that of $\Sigma_k$, with the
		maximum absolute value of $\sim1500$, which is an order of magnitude lower than that of $\Sigma_k$.
		
		\begin{figure*}
			\centering
		    \includegraphics[width=1\textwidth]{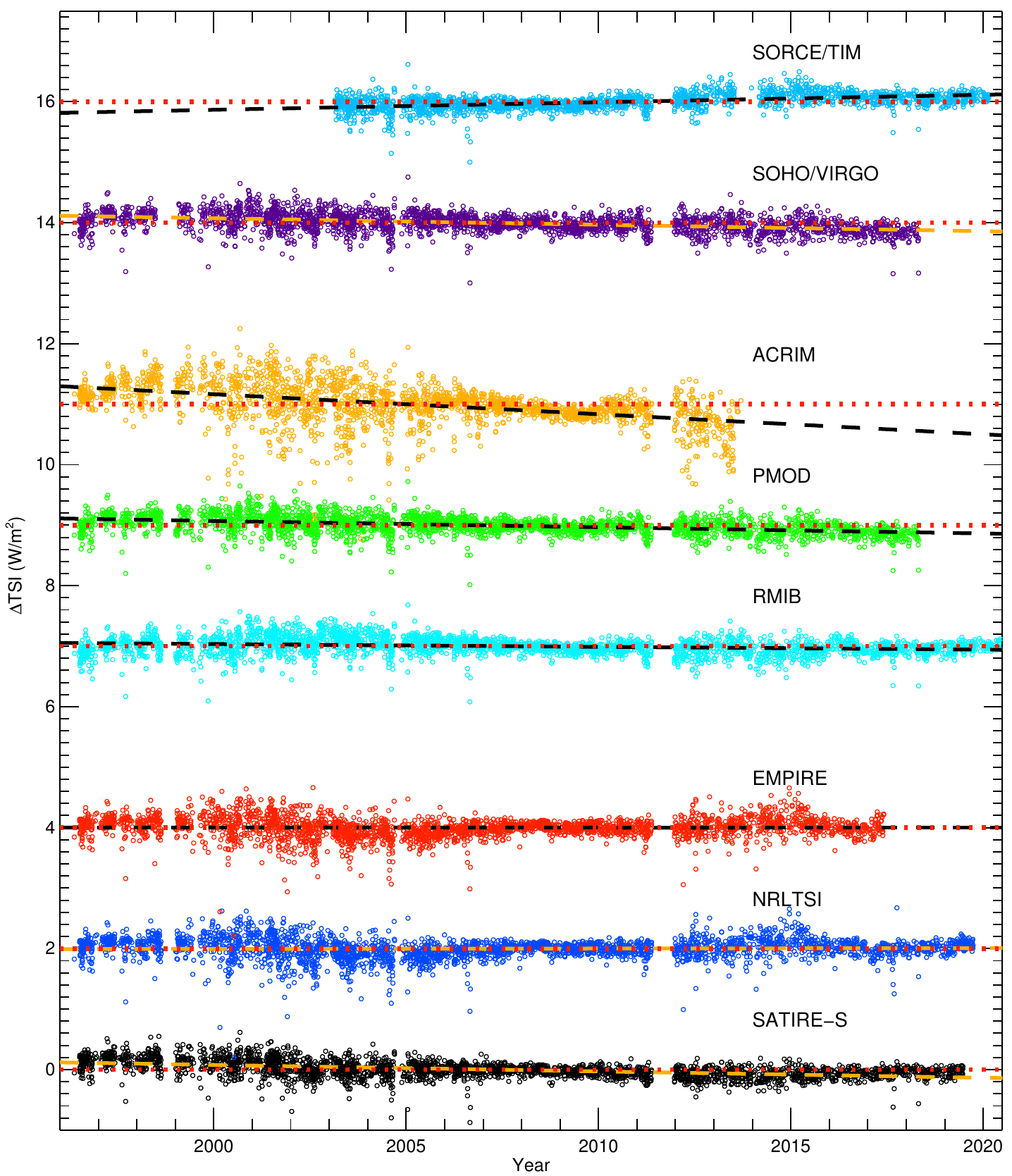}
				\caption{\small Differences 
			 between the various reference TSI datasets (see the legend) and the corresponding TSI reconstructions
				from RP Ca~II~K and blue observations. The residuals have been offset to improve visibility. The dotted red lines denote differences of 0, while the dashed lines are linear fits to the residuals.} 
			\label{fig:tsirp1all}
		\end{figure*}

			The orange line shows the originally derived $\Sigma_k$, $\Sigma_b$, and $\Sigma_r$.
			As was already suggested by Fig. \ref{fig:processedimages_raw}, the contribution of facular regions in  Ca~II~K images is considerably higher than that of sunspots, leading to an in-phase variation of $\Sigma_k$ with the solar cycle, i.e. the Sun in the  Ca~II~K line appears brighter during periods of activity maxima.
            In the red and blue continuum images, where the contrast of faculae is generally lower than that of  sunspots, the balance is rather shaky and
            $\Sigma_b$ and $\Sigma_r$ might vary in anti-phase  with the solar cycle, i.e. the Sun in the specific continuum intervals might appear marginally darker during periods of activity maxima. 
			This is in agreement with the result of \cite{preminger_activity-brightness_2011}.
			However, as discussed below and as also argued by \cite{peck_photometric_2015}, this result turns to be extremely sensitive to the image processing, in particular, to the accuracy of the determination of the CLV.

		In the following we assess the potential uncertainties in the derived photometric sum indices.
			We use two different approaches to identify the QS level on the contrast images and to evaluate its variation with time. 
			This allows us to test whether the processing had introduced any bias in the QS level.
			Here, we focus on the continuum data since such a test on  Ca~II~K observations has already been performed by \citet{chatzistergos_analysis_2017} and \citet{chatzistergos_analysis_2018} using synthetic images. 
		
		Both methods make the underlying assumption that the distribution of the image contrast values is Gaussian, with active regions contributing to the wings of the distribution, but they take different approaches to identify the mean level of the QS regions.
		The first method was proposed by \cite{nesme-ribes_fractal_1996} and was also used by \cite{chatzistergos_analysis_2019,chatzistergos_analysis_2020}.
		An iterative process is used to estimate the level of the QS.
		Initially, the mean contrast, $\bar{C}$, and the standard deviation, $\sigma$, of the contrast values within the disc are computed. 
		Then the mean and the standard deviation are re-calculated iteratively, each time leaving out
		regions with values outside the interval $\bar{C}\pm k \sigma$, where $k$ is a constant.
		The value of $k$ is taken between 0.1 and 3.0, and using a bisector approach, we search for the lowest value of the mean contrast among all $k$ values.
		This lowest mean contrast value is then adopted as the one representing the QS.

		The second approach is an adaptation of the method used by \cite{preminger_photometric_2002} to analyse the potential bias in the SF data.
		For this, 
			we compute the histogram of contrast values for each RP red and blue image, where
			the contrast values are grouped into bins of $10^{-5}$ width. 
			A consistent image processing would return profiles centred at a value constant over the solar cycle. 
			Following \cite{preminger_photometric_2002}, we fit two Gaussians to the histogram of contrast values for each processed image, one for the central part, representing the QS regions, and one for the wing representing the bright regions. 
			To confine the central part, we iteratively reduce the considered interval and fit a Gaussian function, searching for the contrast range for which the lowest $\chi^2$ of the Gaussian fit is achieved.
			The location of the peak of the central Gaussian is taken to represent the mean level of the QS regions.
			The residual between the distribution and the central Gaussian in the bright wing is then fit with another Gaussian function.
			\cite{preminger_photometric_2002} performed the fit on 7 annuli equally spaced in $\mu$ rather than the whole disc.
			They also did not iterate.
			Despite this modification, we will still refer to this approach as the one by \cite{preminger_photometric_2002}.

			As an example, Fig.~\ref{fig:gaussfit} 
			shows the histograms of the RP red and blue images presented in Fig.~\ref{fig:processedimages_raw} 
			along with the evolution of the QS level computed with the two approaches mentioned above for all continuum images analysed in this study.
			Both approaches indicate a weak although non-negligible variation of the QS level in anti-phase with the solar cycle. 
			The QS level derived with the method of \cite{preminger_photometric_2002} varies by more than when derived with the method of \cite{nesme-ribes_fractal_1996}.
			The mean and the standard deviation of the QS level obtained with the method of \cite{preminger_photometric_2002} are $-1\pm4.9\times10^{-4}$ and $-2.5\pm3.8\times10^{-4}$ for the RP red and blue images, respectively.
			The mean and the standard deviation of the QS level determined with the method of \cite{nesme-ribes_fractal_1996} are $-0.7\pm0.3\times10^{-4}$ and $-1.2\pm6.5\times10^{-4}$ for the RP red and blue images, respectively.
			These findings suggest that the QS level in the RP red and blue images calculated with both approaches is affected by the solar activity.
			In particular, the QS level is underestimated during activity maxima.
		Around activity minima, both methods return relatively accurate results. 
		The variation is stronger in the RP blue than in the RP red images. 
		We found no dependence of the bias on the heliocentric angle. 
	However, when applied on narrow annuli, the results are less accurate for both methods.
		The reason is the significantly reduced statistics compared to the case when the entire disc is considered. 
		It is worth noting that the results obtained with the method by \cite{nesme-ribes_fractal_1996} are  qualitatively similar to those obtained by \cite{chatzistergos_analysis_2017} from the analysis of Ca~II~K data, for which we found a bias of $\sim1\pm1\times10^{-4}$ with a minute anti-phase relation to the solar cycle.

		In \cite{chatzistergos_analysis_2017} we showed that the accuracy of recovering the pixel-by-pixel contrast values in synthetic Ca~II~K images was on average $10^{-3}$ (the RMS error reached up to $3\times10^{-3}$). We also showed that our approach worked better than other published methods.
			This pixel-by-pixel error is one order of magnitude higher than the QS bias reported here with both methods.
			These processing errors can affect the determination of the QS level.
			To estimate the sensitivity of the applied methods
			on the pixel-by-pixel errors of our processing, we performed a Monte Carlo simulation.
			In particular, we used all RP contrast images and imposed a random offset on contrast values in each pixel. We then recomputed the QS level with the two considered methods.
			We have done this for 100 different sets of uniformly distributed random contrast offsets within the range -3--3$\times10^{-3}$.
			The distribution of the differences between the QS level computed from the offset and from the original images gives us an estimate of the uncertainty of the QS bias estimation due to processing errors.
			These distributions for annual bins are shown in Fig.~\ref{fig:gaussfit}.
			We find the uncertainty due to the pixel-by-pixel errors to typically lie below $10^{-4}$ and $3\times10^{-5}$ for the methods by \cite{preminger_photometric_2002} and \cite{nesme-ribes_fractal_1996}, respectively.
			This suggests that the estimation of the QS bias is not strongly affected by the pixel-by-pixel processing errors of our method.
			The uncertainty with both methods is comparable, although the effect is slightly stronger when using the method by \cite{preminger_photometric_2002}. 
			The derived QS bias for RP blue data is higher over the years 1996 and 1997, thus rendering our results over these two years rather uncertain.
			We note, however, that this uncertainty estimate of the QS bias is only valid for images processed with our method and it is expected to be higher when using other published processing methods.

		The results presented in Fig.~\ref{fig:gaussfit} 
		suggest that to accurately determine the photometric sums $\Sigma_b$ and $\Sigma_r$,
		we need to account for the bias in the level of the QS definition.
		Therefore, we repeated the computation of the  $\Sigma$ series after the appropriate adjustment of the QS level in each image, using both the methods by \cite{preminger_photometric_2002} and \cite{nesme-ribes_fractal_1996}.
		These adjusted photometric sum series are shown in Figure~\ref{fig:phsumrp1pmod}.
		The RMS difference between the $\Sigma$ series adjusted  with the method by \cite{nesme-ribes_fractal_1996} and the original $\Sigma$ series is 115 for both the red and the blue data.
		Using the method by \cite{preminger_photometric_2002} the RMS differences are 294 and 384 for the red and blue observations, respectively.
		For comparison, the corresponding RMS difference for the Ca~II~K data when using the method by \cite{nesme-ribes_fractal_1996} is 146.
		The adjustment to account for the bias of the $\Sigma_k$ series has barely any effect on the series since the adjustment is two orders of magnitude lower than the $\Sigma$ values. 
		However, for the blue and red observations, the correction is comparable to the $\Sigma$ values.

		After the correction, the temporal profile of $\Sigma_b$ and $\Sigma_r$ series changes both in terms of the amplitude and the shape (see Fig.~\ref{fig:phsumrp1pmod}).
		In particular, with the QS level corrected with the method by \citet[][red curves]{nesme-ribes_fractal_1996} the profile of $\Sigma_b$ becomes essentially flat, with only a tiny decrease
		over the maximum of solar cycle 24, whereas the variation of the $\Sigma_r$ remains in anti-phase with the solar cycle.
		After the correction of the QS level with the method by \cite{preminger_photometric_2002} the variation of $\Sigma_b$ turns to be in-phase with the solar cycle, while the profile of $\Sigma_r$ becomes essentially flat.
		However, while the adjusted $\Sigma_r$ and $\Sigma_b$ are, on average, close to zero at minima when the method by \cite{nesme-ribes_fractal_1996} is used, the corresponding
		values for $\Sigma_b$ with the corrections following \cite{preminger_photometric_2002} are roughly around $100$.
		This points to a potential over-correction with the method by \cite{preminger_photometric_2002}.
		We thus conclude, that the uncertainty in the definition of the QS bias, independently of the method used, as well as the potentially remaining dependence of the QS bias on solar activity
		render the phase of the variation of $\Sigma_r$ and $\Sigma_b$ significantly uncertain.
		We consider the method by \cite{nesme-ribes_fractal_1996} to return more accurate results than the method by \cite{preminger_photometric_2002} as applied here.
		In the following we will use the $\Sigma$ series adjusted for the QS bias with the method by \cite{nesme-ribes_fractal_1996} being the intermediate case found here.

		We note that \cite{preminger_photometric_2002} also reported a variation of $\Sigma_r$ derived from SF red continuum images similar to what we obtain here.
		However, they argued that the offset was too small to have any effect on the solar-cycle phase of the $\Sigma_r$ variability.
		Specifically, they found a mean bias in the QS level during an activity maximum of only $\sim5\times10^{-4}$, although
		this offset doubled towards the limb.
		They argued that this bias introduced a mean offset in $\Sigma_r$ of 120 and -13 for 1989 and 1996, corresponding to activity maximum and minimum periods, respectively.
		These values are consistent in amplitude with the ones derived here. 

		\cite{puiu_modeling_2019} analysed the accuracy of $\Sigma$ evaluation by processing a small sub-sample of RPS observations with methods different than those employed in this study (see Sect.~\ref{sec:discussion} for a brief description of their method). In particular, 
		they studied the effect of various parameters of the processing on their results and especially on the photometric sums.
		Their final adopted parameters returned contrast images favouring the variation of $\Sigma_r$ in anti-phase with the solar cycle.
		However, their analysis showed
		that generally the image processing can result in a reversal of the phase of the relation between $\Sigma_r$ and the solar cycle.

	\section{Long-term trends in reconstructed TSI}
		\label{sec:tsi}
		
		Using the computed photometric sums and Eq.~\ref{eq:tsi}, we now reconstruct the TSI.
		The free parameters of the multiple linear regression model are determined by fitting various combinations of the photometric sums, $\Sigma$, to the PMOD TSI composite.
		We consider the $\Sigma$ series derived from the following 5 pairs of observations: RP Ca~II~K and RP red, RP Ca~II~K and RP blue, RPS Ca~II~K and RPS red, as well as SF Ca~II~K and RP red and SF Ca~II~K and RP blue.
		Figure~\ref{fig:tsirp1pmod} 
		shows the reconstructed TSI along with the PMOD TSI composite.
		Table \ref{tab:tsirmsdif2} summarises the parameters of the fit, along with the RMS difference, and linear correlation coefficient, $R$, for each reconstruction.
		We found that the  best TSI reconstruction is achieved with RP Ca~II~K and RP blue observations, with a RMS difference and $R$ to the PMOD TSI series of 0.15 Wm$^{-2}$ and 0.96, respectively. 
		The TSI reconstructions from RP red observations exhibit a slightly higher scatter than those derived from RP blue data. 
        The quality of the TSI reconstructions is similar to the previous combinations when RPS or SF Ca~II~K data are used.
		The PMOD TSI shows a somewhat stronger declining trend than the reconstruction.
		This is discussed in more detail below.
		
		The empirical TSI reconstructions, based on the regression of solar activity indices to the TSI measurements, are obviously subject to the uncertainties in those measurements.
		Therefore,
		we have done various TSI reconstructions by varying the TSI reference series used for the regression.
		Table~\ref{tab:tsirmsdif2} 
		lists the parameters of the fit, along with the RMS difference, and $R$ for each reconstructions, whereas  
		Fig.~\ref{fig:tsirp1all} 
		shows the residual between each considered TSI reference record and the corresponding reconstruction from RP Ca~II~K and RP blue observations. 
		
		We generally found a good agreement between our TSI reconstructions and the various published TSI series (see Sect. \ref{sec:tsirefseries}). 
		Most reconstructions show the same characteristics.
		In particular, as in the case of PMOD, a slightly declining trend is seen in the residuals with time.
		Exceptions are the reconstructions that use the SORCE/TIM data as reference, which display a slightly increasing trend of $1.3\times10^{-2}$ Wm$^{-2}$y$^{-1}$.
		This might partly be because the SORCE/TIM data do not cover the maximum of cycle 23.
		However, we note that \cite{dewitte_total_2016} and \cite{woods_decoupling_2018} reported an increasing drift of $3.43\times10^{-2}$ Wm$^{-2}$y$^{-1}$ in the TSI values from SORCE/TIM over the same period.
		We also found that our TSI reconstruction with the ACRIM composite as the reference displays the strongest trend of residuals to the measurements. This is most likely due to issues with the  ACRIM composite series.
		We also computed the R and RMS differences between our reconstructed TSI and the various reference series after detrending our series. 
		For detrending 
		we used the residual trend listed in Table 2. We find the RMS differences to generally decrease slightly in all cases, however qualitatively we report the same findings.

For a more quantitative comparison,
		we have performed a linear fit to the reconstructed TSI over each two subsequent solar activity minima (that is 1996 vs. 2008 and 2008 vs. 2019).
		For this, we used 12-month intervals centered at the minimum of each of the 3 cycles.
		The dates of the minima were taken as 01 August 1996, 01 December 2008, and 01 October 2019 \citep{hathaway_solar_2015}.  
		We note that at the time of writing this manuscript there is no official starting date for solar cycle 25 and hence the value used here might change somewhat, although probably not much \citep[see e.g.][]{alterman_helium_2020}.
		Therefore, we use the last year for which the data are currently available for the last minimum.

\begin{figure*}
	\centering
	\includegraphics[width=1\textwidth]{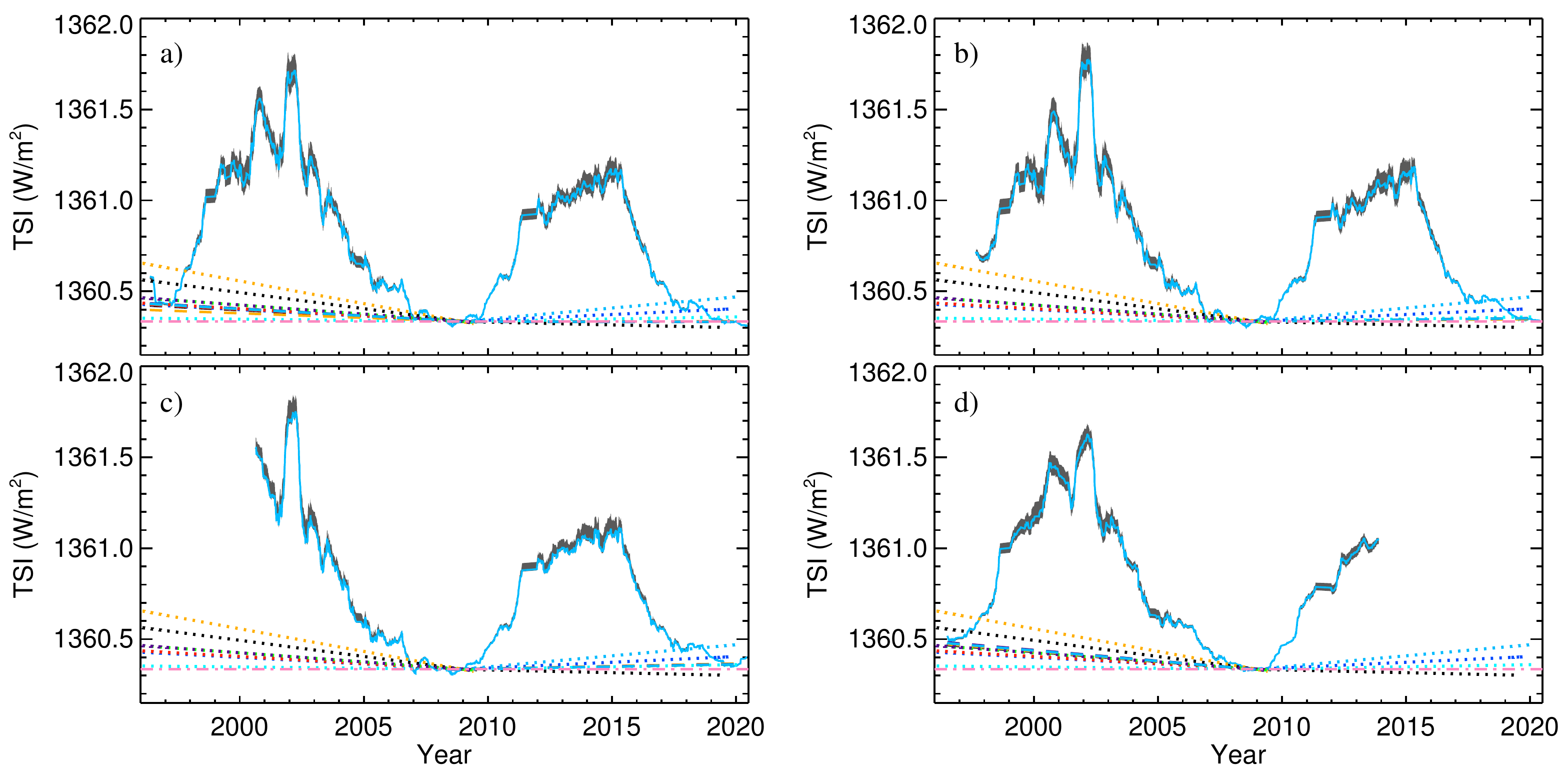}
		\caption{\small TSI reconstructed with various reference series using the following images: RP Ca~II~K and RP blue (a); RP Ca~II~K and RP red (b); RPS CA~II~K and RPS red (c); and SF Ca~II~K and RP blue (d).
		Colours are the same as in Fig. \ref{fig:tsirp1all}.
		The solid line is 81-day running mean values for the series with PMOD as the reference, while the grey shaded surface shows the range covered by the reconstructions with all other TSI reference series. The dashed and dotted lines are linear fits to the reconstructed and reference TSI values, respectively, during subsequent activity minima.  
		All series were offset to match their mean value within a 12-month interval centered at the minimum of 2008 to that of the PMOD TSI composite. The value of the PMOD TSI composite over this period is marked with the horizontal dash-dotted pink line. The various dashed and dotted lines extend only to the intervals for which their respective data cover the complete two surrounding minima.} 
	\label{fig:tsirp1all_entire}
\end{figure*}

		Figure \ref{fig:tsirp1all_entire} 
		shows the reconstructions for the various TSI measurement records employed as the reference, along with the linear fits to the TSI values during the activity minima.
		For ease of comparison, we offset the reconstructed series so to match their mean values over a 12-month interval centred at the minimum of 2008 to that of the PMOD TSI series.
		While the various reconstructions differ slightly in the magnitude of the solar cycle variation, they agree remarkably well during activity minima, showing essentially the same minimum-to-minimum changes.
		Thus, the choice of the reference TSI dataset is not decisive for the long-term trend in the reconstructed record.
		We also note that the long-term trends of the TSI series derived from the  PMOD and RMIB TSI composite data employed as the references are remarkably close to each other, with differences being less than $2.1\times10^{-4}$ Wm$^{-2}$y$^{-1}$. However, this is mainly due to the similarity of the PMOD and RMIB TSI composites over the selected periods.

		The TSI reconstruction based on RP Ca~II~K and RP blue images that uses the PMOD data as the reference shows a long-term trend of -7.8$^{+4.9}_{-0.8}\times10^{-3}$ and -0.1$^{+0.25}_{-0.02}\times10^{-3}$ Wm$^{-2}$y$^{-1}$  
		between the minima 1996/2008 and 2008/2019, respectively, where the uncertainty ranges account for the spread in results derived using all other published TSI data as the reference. 
		The uncertainty of the fit is roughly $10^{-3}$ Wm$^{-2}$y$^{-1}$ and is lower than the uncertainty due to using different TSI reference series. The declining trend over the 1996/2008 minima is statistically significant based on a Student's t-test, but this is not the case for the declining trend over the 2008/2019 minima. 
		However, using different input data returns slightly different values for the long-term trend. 
		For example, for the change between 1996 and 2008 we found $-11.9^{+1.2}_{-0.06}\times10^{-3}$ Wm$^{-2}$y$^{-1}$ when using SF Ca~II~K and RP blue observations, which is slightly higher than obtained from RP Ca~II~K and blue data. 
		Between the 2008 and 2019 minima, the trend marginally changes sign and is $2.6^{+0.3}_{-0.1}\times10^{-3}$ Wm$^{-2}$y$^{-1}$ with RPS Ca~II~K and red observations, while it becomes $1.6^{+0.7}_{-0.1}\times10^{-3}$ 
		Wm$^{-2}$y$^{-1}$ with RP Ca~II~K and red observations.
		Overall, our reconstructions suggest a weak decline in TSI from 1996 to 2008, and essentially no trend 
		from 2008 to 2019.
		
Finally, for completeness, we have also compared our reconstructions to the widely used models, namely SATIRE-S,
NRLTSI and EMPIRE (see Sect.~\ref{sec:tsirefseries} for the description of the datasets and Table~2 and Figs. 5 and 6 for the results).
Thereby, for an unbiased comparison, we also used the corresponding series as the reference.
The lowest RMS difference and highest $R$ are obtained in comparison to SATIRE-S and  EMPIRE reconstructions.
The slope of the residual is somewhat closer to zero when the two empirical reconstructions, EMPIRE and NRLTSI, are used as reference. These models are similar in concept to ours, i.e. they are also based on linear regressions of the solar activity indices to the measured TSI, except that they use disc-integrated proxies.
The slope of the residual between our reconstruction and SATIRE-S is closer to that obtained for most reconstructions using observational records as the reference (except SORCE/TIM).
All differences are, however, rather small.

		\begin{table*}
			\caption{Results of reconstructing TSI.}
			\label{tab:tsirmsdif2}     
			\centering                      
				\begin{tabular}{l*{10}{c}}       
					\hline\hline                
					$\Sigma_k$ &$\Sigma_{r,b}$& TSI & $a$&$b$&$c$&RMS & $R$ & Slope &\multicolumn{2}{c}{Trend}\\
					&           &   &&    &   & &    &&  1996--2008 & 2008--2019\\		    	        &           &   &Wm$^{-2}$&    & $10^{-2}$  &Wm$^{-2}$ &    &$10^{-2}$Wm$^{-2}$y$^{-1}$&\multicolumn{2}{c}{$10^{-3}$Wm$^{-2}$y$^{-1}$} \\
					\hline   
RP & RP Blue & PMOD      & 1360.20$\pm$0.04 & 1.20$\pm$0.06 & 5.5$\pm$0.2 & 0.15 & 0.96 & -1.03 & -7.8$\pm$1.1 & -0.10$\pm$0.87\\
RP & RP Blue & ACRIM     & 1360.72$\pm$0.04 & 0.75$\pm$0.07 & 5.5$\pm$0.3 & 0.34 & 0.79 & -3.31 & -5.0$\pm$0.8 &  0.15$\pm$0.67\\
RP & RP Blue & RMIB      & 1362.74$\pm$0.03 & 1.18$\pm$0.06 & 5.4$\pm$0.2 & 0.14 & 0.96 & -0.46 & -7.7$\pm$1.1 & -0.10$\pm$0.86\\
RP & RP Blue & SORCE/TIM & 1360.41$\pm$0.04 & 1.19$\pm$0.09 & 5.2$\pm$0.4 & 0.13 & 0.94 &  1.25 & -7.7$\pm$1.1 & -0.12$\pm$0.85\\
RP & RP Blue & SOHO/VIRGO& 1360.19$\pm$0.04 & 1.21$\pm$0.06 & 5.6$\pm$0.2 & 0.16 & 0.95 & -1.07 & -7.9$\pm$1.1 & -0.10$\pm$0.88\\
RP & RP Blue & EMPIRE    & 1360.34$\pm$0.04 & 1.11$\pm$0.06 & 5.9$\pm$0.2 & 0.17 & 0.94 & -0.02 & -7.3$\pm$1.0 & -0.01$\pm$0.84\\
RP & RP Blue & NRLTSI    & 1360.40$\pm$0.03 & 1.13$\pm$0.06 & 5.5$\pm$0.2 & 0.17 & 0.94 &  0.06 & -7.3$\pm$1.0 & -0.06$\pm$0.83\\
RP & RP Blue & SATIRE-S  & 1360.37$\pm$0.03 & 1.06$\pm$0.06 & 5.4$\pm$0.2 & 0.14 & 0.96 & -1.06 & -6.9$\pm$1.0 & -0.03$\pm$0.80\\
			\hline   
RP & RP Red & PMOD      & 1360.18$\pm$0.04 & 1.67$\pm$0.09 & 6.7$\pm$0.3 & 0.18 & 0.94 & -1.32 & - &  1.65$\pm$0.74\\
RP & RP Red & ACRIM     & 1360.67$\pm$0.04 & 1.03$\pm$0.11 & 6.4$\pm$0.3 & 0.35 & 0.79 & -3.43 & - &  1.19$\pm$0.62\\
RP & RP Red & RMIB      & 1362.74$\pm$0.03 & 1.64$\pm$0.09 & 6.6$\pm$0.3 & 0.17 & 0.94 & -0.87 & - &  1.63$\pm$0.72\\
RP & RP Red & SORCE/TIM & 1360.39$\pm$0.04 & 1.71$\pm$0.13 & 6.5$\pm$0.4 & 0.13 & 0.94 &  1.11 & - &  1.67$\pm$0.73\\
RP & RP Red & SOHO/VIRGO& 1360.17$\pm$0.04 & 1.68$\pm$0.09 & 6.8$\pm$0.3 & 0.18 & 0.93 & -1.34 & - &  1.66$\pm$0.74\\
RP & RP Red & EMPIRE    & 1360.32$\pm$0.04 & 1.56$\pm$0.09 & 7.1$\pm$0.3 & 0.19 & 0.93 &  0.02 & - &  1.60$\pm$0.74\\
RP & RP Red & NRLTSI    & 1360.38$\pm$0.04 & 1.59$\pm$0.09 & 6.7$\pm$0.3 & 0.18 & 0.93 &  0.02 & - &  1.59$\pm$0.72\\
RP & RP Red & SATIRE-S  & 1360.34$\pm$0.04 & 1.47$\pm$0.09 & 6.6$\pm$0.3 & 0.16 & 0.95 & -1.10 & - &  1.50$\pm$0.69\\
			\hline   
RPS & RPS Red & PMOD      & 1360.16$\pm$0.04 & 1.70$\pm$0.10 & 7.4$\pm$0.3 & 0.17 & 0.94 & -1.32 & - &  2.57$\pm$1.32\\
RPS & RPS Red & ACRIM     & 1360.63$\pm$0.05 & 0.99$\pm$0.12 & 6.8$\pm$0.4 & 0.33 & 0.81 & -3.17 & - &  2.87$\pm$0.87\\
RPS & RPS Red & RMIB      & 1362.71$\pm$0.04 & 1.66$\pm$0.10 & 7.4$\pm$0.3 & 0.17 & 0.94 & -1.08 & - &  2.56$\pm$1.29\\
RPS & RPS Red & SORCE/TIM & 1360.37$\pm$0.04 & 1.70$\pm$0.13 & 7.4$\pm$0.5 & 0.14 & 0.93 &  1.00 & - &  2.53$\pm$1.32\\
RPS & RPS Red & SOHO/VIRGO& 1360.16$\pm$0.04 & 1.71$\pm$0.10 & 7.5$\pm$0.3 & 0.18 & 0.94 & -1.32 & - &  2.58$\pm$1.33\\
RPS & RPS Red & EMPIRE    & 1360.30$\pm$0.04 & 1.57$\pm$0.10 & 7.7$\pm$0.3 & 0.17 & 0.94 &  0.49 & - &  2.85$\pm$1.25\\
RPS & RPS Red & NRLTSI    & 1360.36$\pm$0.04 & 1.60$\pm$0.10 & 7.3$\pm$0.3 & 0.17 & 0.94 &  0.33 & - &  2.61$\pm$1.25\\
RPS & RPS Red & SATIRE-S  & 1360.32$\pm$0.04 & 1.50$\pm$0.10 & 7.2$\pm$0.3 & 0.14 & 0.95 & -0.91 & - &  2.60$\pm$1.18\\
\hline
SF & RP Blue & PMOD      & 1360.19$\pm$0.05 & 0.93$\pm$0.08 & 36.2$\pm$2.2 & 0.17 & 0.94 & -0.49 & -11.9$\pm$1.3 &  -\\
SF & RP Blue & ACRIM     & 1360.69$\pm$0.05 & 0.79$\pm$0.09 & 37.3$\pm$2.3 & 0.32 & 0.83 & -3.03 & -11.1$\pm$1.1 &  -\\
SF & RP Blue & RMIB      & 1362.72$\pm$0.05 & 0.92$\pm$0.08 & 36.0$\pm$2.2 & 0.17 & 0.94 &  0.07 & -11.8$\pm$1.2 &  -\\
SF & RP Blue & SORCE/TIM & 1360.34$\pm$0.08 & 0.93$\pm$0.14 & 34.8$\pm$4.6 & 0.14 & 0.92 &  1.98 & -11.6$\pm$1.2 &  -\\
SF & RP Blue & SOHO/VIRGO& 1360.18$\pm$0.05 & 0.94$\pm$0.08 & 36.4$\pm$2.2 & 0.18 & 0.94 & -0.45 & -11.9$\pm$1.3 &  -\\
SF & RP Blue & EMPIRE    & 1360.31$\pm$0.05 & 0.81$\pm$0.08 & 37.9$\pm$2.2 & 0.20 & 0.93 & -0.07 & -11.3$\pm$1.1 &  -\\
SF & RP Blue & NRLTSI    & 1360.38$\pm$0.05 & 0.83$\pm$0.08 & 35.1$\pm$2.2 & 0.20 & 0.92 & -0.02 & -11.0$\pm$1.1 &  -\\
SF & RP Blue & SATIRE-S  & 1360.37$\pm$0.05 & 0.78$\pm$0.08 & 35.4$\pm$2.2 & 0.16 & 0.94 & -1.15 & -10.7$\pm$1.1 &  -\\
			\hline   
SF & RP Red & PMOD      & 1360.19$\pm$0.06 & 1.36$\pm$0.12 & 42.2$\pm$2.5 & 0.17 & 0.94 & -1.03 & - &  -\\
SF & RP Red & ACRIM     & 1360.65$\pm$0.06 & 1.04$\pm$0.12 & 43.0$\pm$2.6 & 0.34 & 0.82 & -3.35 & - &  -\\
SF & RP Red & RMIB      & 1362.75$\pm$0.06 & 1.35$\pm$0.12 & 41.1$\pm$2.5 & 0.17 & 0.95 & -0.75 & - &  -\\
SF & RP Red & SORCE/TIM & 1360.34$\pm$0.08 & 1.37$\pm$0.19 & 40.5$\pm$4.7 & 0.13 & 0.94 &  1.93 & - &  -\\
SF & RP Red & SOHO/VIRGO& 1360.19$\pm$0.06 & 1.27$\pm$0.12 & 41.9$\pm$2.5 & 0.19 & 0.93 & -0.90 & - &  -\\
SF & RP Red & EMPIRE    & 1360.30$\pm$0.06 & 1.18$\pm$0.12 & 43.7$\pm$2.5 & 0.20 & 0.93 & -0.24 & - &  -\\
SF & RP Red & NRLTSI    & 1360.37$\pm$0.06 & 1.13$\pm$0.12 & 40.3$\pm$2.5 & 0.21 & 0.91 & -0.10 & - &  -\\
SF & RP Red & SATIRE-S  & 1360.34$\pm$0.06 & 1.04$\pm$0.12 & 40.8$\pm$2.5 & 0.18 & 0.93 & -1.22 & - &  -\\
					\hline
				\end{tabular}
				\tablefoot{Columns are: Ca~II~K and continuum observations employed to compute $\Sigma_k$, $\Sigma_b$, and $\Sigma_r$, TSI reference series used for the regression, best fit parameters of Eq. \ref{eq:tsi}, RMS difference, linear correlation coefficient $R$, and slope of residual  between the reconstructed TSI series and the reference one, and long-term trend as defined with a linear fit during activity minimum periods.}
		\end{table*}

		\section{Discussion}
		\label{sec:discussion}
		In Sect. \ref{sec:phase}
		we analysed the sensitivity of the $\Sigma$ series to the computation of the QS level in the images when processed with our method.  
		We now consider also the effect of the QS computation on the $\Sigma$ series and the TSI reconstruction when using different processing methods.
		In particular, we consider the methods presented by \cite{brandt_determination_1998}, \cite{walton_processing_1998}, \cite{worden_evolution_1998}, and \cite{puiu_modeling_2019}. 
		All of these studies used full-disc  Ca~II~K observations to reconstruct TSI. 

		Briefly, \cite{brandt_determination_1998} and \cite{walton_processing_1998} derive the QS CLV by considering image pixels grouped into concentric annuli, with the value of the QS within each annulus determined with a threshold in the cumulative histogram of intensity values and the median of intensity values, respectively. 
		Since both of those methods assume that the CLV is perfectly radially symmetric, they are unable to account for some image artefacts that often affect the full-disc observations, such as the linear gradient of intensity values caused by the atmospheric extinction in the RP and RPS data. 
		To account for this, 
		we have added an extra step to perform a 2D linear fit to the contrast images resulting from the processing by \cite{brandt_determination_1998} and \cite{walton_processing_1998}. This fit is then subtracted from the contrast images. 
		The processing by \cite{worden_evolution_1998} incorporates 5th degree polynomial fits along lines in the vertical and horizontal directions, as well as in segments with the orientation of 45$^\circ$ and 135$^\circ$.
		Finally, \cite{puiu_modeling_2019} calculates the background as \cite{walton_processing_1998}, but 
		regions within and outside 0.7$r$ are processed separately. 
		Additionally, active regions are excluded from the QS computation with an iterative procedure that follows the one proposed by \cite{chatzistergos_analysis_2018} and used in this study. 
		As the last steps, \cite{puiu_modeling_2019} applies a linear vertical fit and a running window median filter to account for any residual intensity gradient in RPS data.

		In \cite{chatzistergos_analysis_2017} and \cite{chatzistergos_analysis_2018} we studied the accuracy of
		the methods proposed 
		by \cite{brandt_determination_1998}, \cite{walton_processing_1998}, and \cite{worden_evolution_1998} to remove the limb-darkening on synthetic Ca~II~K data.  
		We showed that these methods performed worse than the ones employed in this study, by introducing artefacts 
		that are subject to the solar cycle due to an inaccurate exclusion of active regions when defining the QS CLV. 
		In particular, 
		when using the methods by \cite{brandt_determination_1998}, \cite{walton_processing_1998}, and \cite{worden_evolution_1998},
		 mean offsets for the QS level were up to 2, 1, and 2 orders of magnitude higher than the one obtained with the method employed here, respectively. Furthermore, 
		\cite{puiu_modeling_2019} compared the accuracy of their processing to ours and also found the latter to be more accurate.
		
\begin{figure*}
			\centering
			\includegraphics[width=1\textwidth]{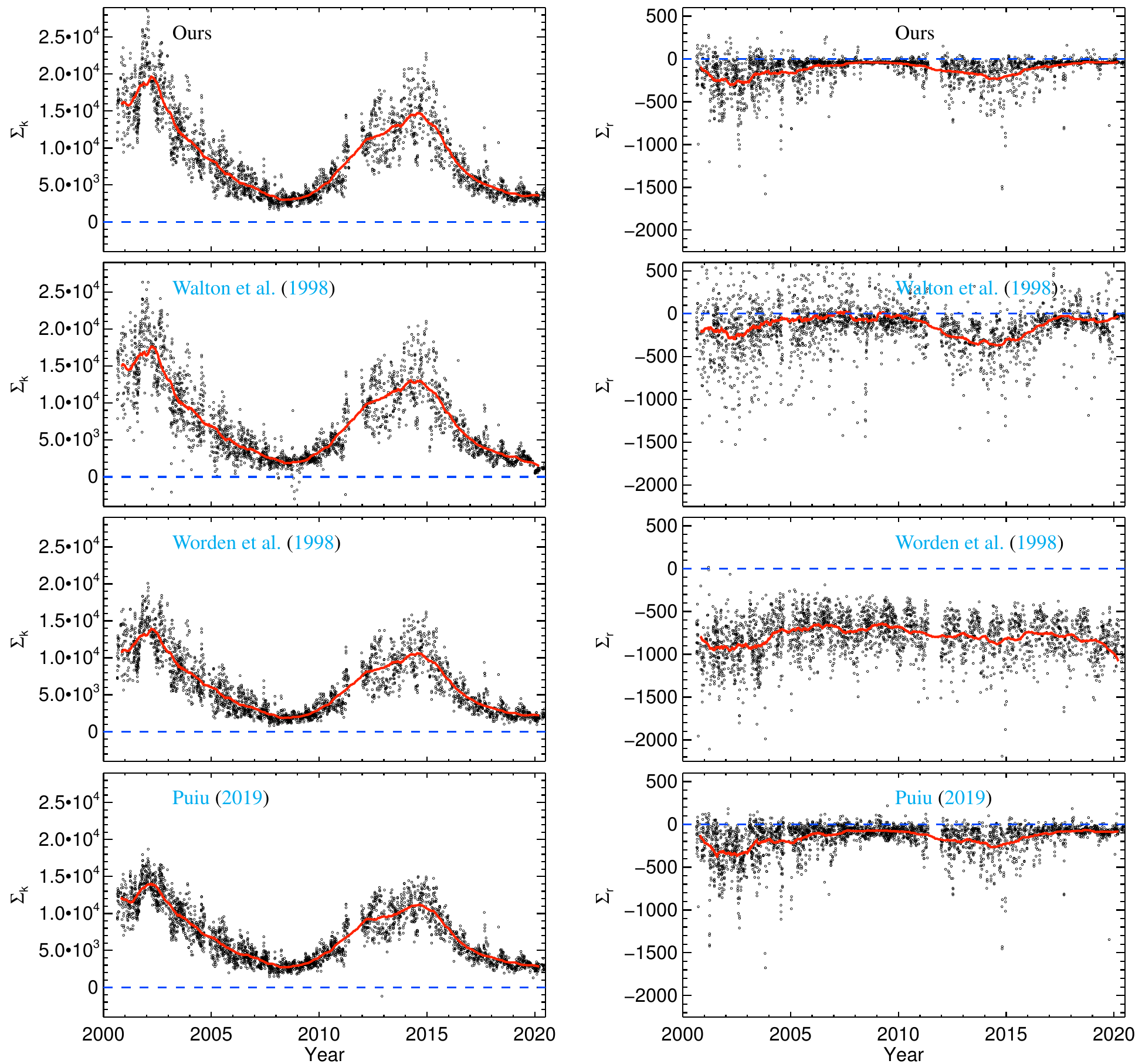}
				\caption{\small Series of photometric sum indices derived from RPS Ca~II~K (left) and RPS red continuum (right) observations which were processed with our method (1st row) and the methods by \citet[][2nd row]{walton_processing_1998}, \citet[][3rd row]{worden_evolution_1998}, and \citet[][4th row]{puiu_modeling_2019}. Daily values are shown in black, while 180-day running averages are shown in red. The dashed blue horizontal line denotes the 0-level of photometric sum.} 
			\label{fig:phsum_otherprocessing}
		\end{figure*}

        It is important to note that these error estimates were limited to Ca~II~K images, for which 
        a significant source of errors is inaccurate accounting for active regions.
        On continuum images, where the disc coverage by active regions is significantly lower than in Ca~II~K images, 
        other methods might fare better. 
        Therefore, to check this, 
        we used all these four methods to process 
        the RPS Ca~II~K and red images, produce photometric sum series and reconstruct the TSI. 
		We restricted this test to the RPS data
		for consistency with the analysis by \cite{puiu_modeling_2019}. 
		Figure \ref{fig:phsum_otherprocessing} 
		(left panels) shows the $\Sigma_k$ series derived from RPS Ca~II~K observations processed with our and three of the earlier methods.
		We do not show the results with the method by \cite{brandt_determination_1998}, because they are very similar to those by \cite{walton_processing_1998}.
		The $\Sigma_k$ series obtained from our method exhibits the highest variability over the solar cycle. 
		This suggests a potential suppression of the contrast of the bright regions by the other processing techniques, which is in agreement with the finding that these methods do not accurately exclude active regions when computing the QS CLV \citep{chatzistergos_analysis_2018}.

		Figure \ref{fig:phsum_otherprocessing} 
		(right panels) shows the $\Sigma_r$ series derived from RPS red images by using our and three of the tested methods.
		All series exhibit a variation in anti-phase with the solar cycle.
		However, the series obtained with the methods by \cite{brandt_determination_1998} and \cite{walton_processing_1998} show high scatter and 
		include many extreme values that are likely due to image artefacts unaccounted for by the image processing. 
		For this reason, we ignored 77 images for which the processing by the method of \cite{walton_processing_1998} produced $\Sigma_r$ values $>$2000.
		We note that these results were obtained by applying the 2D linear fit, which might not accurately remove the linear gradient in the data.
		However, the results without the  additional 2D fit were considerably worse, lacking any clear cycle variation. 
		We also note that the 5th degree polynomial used for the fitting along linear segments in the method by \cite{worden_evolution_1998} fails to reproduce the CLV in RPS red observations.
		This is in agreement with the findings by \cite{puiu_modeling_2019}.  
		We found that the method by \cite{puiu_modeling_2019} results in a distinct anti-phase relation between the $\Sigma_r$ series and the solar cycle.
		The $\Sigma_r$ series computed with the method by \cite{puiu_modeling_2019} shows the best agreement among the various tested methods to that derived with our method, exhibiting a high correlation coefficient of 0.88 (see Table \ref{tab:tsirmsdif_otherprocessing}). This suggests that the method by \cite{puiu_modeling_2019} performs better than the methods by \cite{brandt_determination_1998}, \cite{walton_processing_1998}, and \cite{worden_evolution_1998}.
		The spread of $\Sigma_r$ values during minima is greater when using the method by \cite{puiu_modeling_2019} than with our method.

\begin{figure*}
			\centering
			\includegraphics[width=1\textwidth]{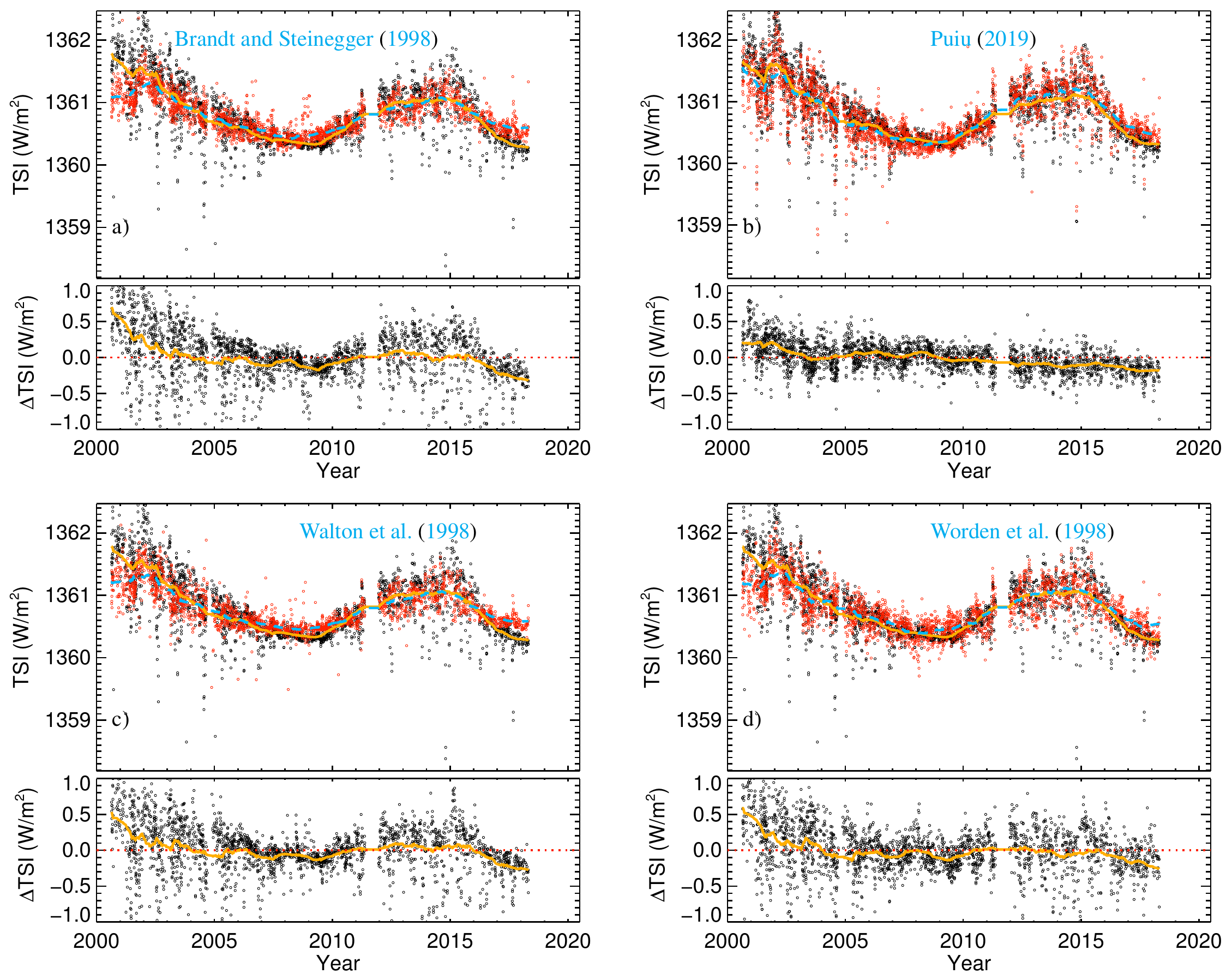}
				\caption{\small Comparison between the PMOD TSI composite series (black circles for daily values and solid orange line for 180-day running mean values) and TSI series reconstructed (red circles for daily values and dashed ciel line for 180-day running mean values) from RPS Ca~II~K and RPS red  observations processed with the methods by \citet[][panel a]{brandt_determination_1998}; \citet[][panel b]{puiu_modeling_2019}; \citet[][panel c]{,walton_processing_1998}; and \citet[][panel d]{worden_evolution_1998}. The PMOD TSI composite was taken as reference for all those reconstructions. Find more details in Section \ref{sec:discussion}. At the bottom of each panel, the corresponding difference between the PMOD TSI and the reconstructed TSI is shown (black circles for daily values and orange line for the 180-day running mean).} 
			\label{fig:tsirp1otherprocessing}
		\end{figure*}

		Table \ref{tab:tsirmsdif_otherprocessing} summarizes the fit parameters (Eq.~\ref{eq:tsi}) when using the $\Sigma$ series derived from all tested methods.  
		Also listed are the RMS difference, $R$, the slope of the residuals to the reference series, as well as the long-term trend between the last solar cycle minima in 2008 and 2019.
		Figure~\ref{fig:tsirp1otherprocessing} 
		shows the TSI variations reconstructed from RPS  Ca~II~K and red data processed with the various tested methods. 
		The TSI reconstruction from RPS data processed with our method is plotted in Fig.~\ref{fig:tsirp1pmod}c.

		TSI reconstructed from the images processed with our method scores better than when applying other processing methods tested here,  followed by the
		method by \cite{puiu_modeling_2019} with
		 slightly worse RMS differences to the PMOD series.
		 		All other processing techniques give worse match to the TSI reference series, with RMS differences increasing to 0.35--0.4 Wm$^{-2}$ instead of 0.17 Wm$^{-2}$ obtained with our method. 
		 Interestingly, the long-term change in TSI between the minima in 2008 and 2019 returned by different methods shows opposite signs.
		Specifically, the trend in TSI determined with the  processing following \cite{puiu_modeling_2019}  and \cite{worden_evolution_1998} is negative (decreasing TSI), and is $-1.5\times10^{-3}$ Wm$^{-2}$y$^{-1}$ and $-10.4\times10^{-3}$ Wm$^{-2}$y$^{-1}$, respectively.
		The latter is also the steepest trend among all methods.
		For the other three methods, the trend is positive (increasing TSI) and ranges from $2\times10^{-5}$ Wm$^{-2}$y$^{-1}$ for the method by \cite{brandt_determination_1998}
		 to $2.6\times10^{-3}$ Wm$^{-2}$y$^{-1}$ for our method.

		These tests suggest that for reliable irradiance reconstructions, accurate processing of the images is crucial.
		Uncertainties of the processing methods might
considerably affect the determined long-term trends in the reconstructed TSI.

		\begin{table*}
			\caption{Results of reconstructing TSI over 2000--2020 from RPS contrast images processed with different methods.}
			\label{tab:tsirmsdif_otherprocessing}     
			\centering    
			\begin{tabular}{l*{9}{c}}       
				\hline\hline                
				Method&	\multicolumn{2}{c}{$R_0$} & $a$&$b$&$c$&RMS & $R$ & Slope &Trend\\
				&$\Sigma_k$&$\Sigma_r$          &(Wm$^{-2}$)&    & $10^{-2}$  &(Wm$^{-2})$ &    &($10^{-2}$Wm$^{-2}$y$^{-1}$)& ($10^{-3}$Wm$^{-2}$y$^{-1}$)\\
				\hline   
				Ours&			-				&        -   &1360.16$\pm$0.04 & 1.70$\pm$0.10 & 7.4$\pm$0.3 & 0.17 & 0.94 & -1.32 &  2.57\\
				\cite{brandt_determination_1998}& 0.95&-0.05&1359.82$\pm$0.07 & 0.01$\pm$0.02 & 5.4$\pm$0.4 & 0.40 & 0.61 & -1.82 & 0.02\\
				\cite{walton_processing_1998}	& 0.98&0.21 &1360.32$\pm$0.04 & 0.23$\pm$0.05 & 4.8$\pm$0.3 & 0.37 & 0.68 & -1.29 &  1.3\\
				\cite{worden_evolution_1998}	& 0.99&0.51 &1360.77$\pm$0.07 & 0.58$\pm$0.07 & 7.1$\pm$0.4 & 0.35 & 0.72 & -1.59 & -10.4\\
				\cite{puiu_modeling_2019}		& 0.97&0.88 &1360.05$\pm$0.04 & 1.33$\pm$0.09 & 10.6$\pm$0.4 & 0.23 & 0.89 & -1.79 & -1.5\\
				\hline
			\end{tabular}
			\tablefoot{Columns are: processing technique applied on the RPS Ca~II~K and RPS red continuum data to compensate the CLV, linear correlation coefficient $R_0$ between the $\Sigma_k$ and $\Sigma_r$ series computed with our processing and those from the various tested methods, the best fit parameters of Eq.~\ref{eq:tsi}, RMS difference, linear correlation coefficient $R$, and slope of residual  between the reconstructed TSI series and the PMOD TSI composite, and long-term trend as defined with a linear fit during the activity minimum period of 2008--2019.}
		\end{table*}

		\section{Summary}
		\label{sec:conclusions}
		
		Solar irradiance has been measured from space for over four decades now. Despite significant progress in our understanding of the irradiance variability, its long-term trend is still under debate.
		To address this question, we have reconstructed the TSI over the period 1996--2020 covering the last two solar cycles.
		
		We have used an empirical model which computes the irradiance changes through a linear regression of solar activity indices.
		We have followed the approach by \cite{chapman_modeling_2013} who used the so-called solar photometric sums as input proxies.
		The photometric sums were produced from full-disc Ca~II~K, red, and blue continuum observations acquired with the Precision Solar Photometric Telescope at the INAF Observatory of Rome and Ca~II~K observations taken with the Cartesian Full-Disk Telescope 2 at the San Fernando Observatory.
		
		We find a good agreement between our reconstructed TSI series and the direct TSI measurements by SOHO/VIRGO and SORCE/TIM, as well as the RMIB and PMOD TSI composites.
		The agreement is worse with the ACRIM TSI composite.
		We also find a good agreement with the empirical NRLTSI and EMPIRE models, as well as with the semi-empirical SATIRE-S model, whereby the agreement with SATIRE-S and EMPIRE is somewhat better.
		
		We find a weak declining trend of -7.8$^{+4.9}_{-0.8}\times10^{-3}$ Wm$^{-2}$y$^{-1}$ between the 1996 and 2008 minima.
		Between the minima in 2008 and 2019, the trend is -0.1$^{+0.25}_{-0.02}\times10^{-3}$ Wm$^{-2}$y$^{-1}$, i.e. the TSI might have decreased slightly although this change is within the error-bars.

		Furthermore, measurements from SORCE/SIM 
		\citep{harder_trends_2009} and from SOHO/VIRGO 
		\citep{wehrli_correlation_2013} suggested conflicting solar-cycle trends in the irradiance variations in the visible part of the spectrum.
        Based on the analysis of photometric sum series from SF red and blue observations, \cite{preminger_photometric_2002} argued that the irradiance variations in the visible are in anti-phase with the solar cycle, in agreement with the SORCE/SIM measurements.
	    Here we used RP observations in the blue and red continuum to produce series of photometric sums and study the phase of their variations. 
        By carefully analysing the accuracy of the image processing applied to the data, we found a small bias in the QS level of RP red and blue continuum observations of the order of 5$\times10^{-4}$ in anti-phase with the solar cycle.
		We argue that this bias renders the photometric sum indices rather uncertain, such that the correlation to solar activity can change from being negative to marginally positive after the correction. The amplitude of the variations in the visible is very low and is  below the uncertainties, which prevents an assessment of the phase of the variations. 
		Results obtained with alternative techniques are even more uncertain.

		This approach to irradiance reconstruction is unfortunately limited to a more recent period only, due to the lack of narrow-band continuum observations before the 1980's.
		Further modifications to the method, e.g.  accounting for sunspots using alternative data, such as white light observations \citep[e.g.][]{willis_greenwich_2016,pal_solar_2020}, could allow irradiance reconstructions from 
        properly calibrated historical  Ca~II~K observations, e.g. those from Meudon \citep{malherbe_new_2019} or Kodaikanal \citep{chatzistergos_delving_2019} observatories, to extend the reconstruction back to 1892 \citep{chatzistergos_analysis_2020}.

		\begin{acknowledgements}
			We thank the observers at the Rome and San Fernando sites for all their work in carrying out the observing programs. 
			We also thank Greg Kopp and the anonymous referee for their valuable comments and suggestions that improved this manuscript.
			T. C. acknowledges funding from the European Union's Horizon 2020 research and Innovation program under grant agreement No 824135 (SOLARNET).
			This work was supported by the Italian MIUR-PRIN grant 2017 ''Circumterrestrial Environment: Impact of Sun--Earth Interaction''
			and by the German Federal Ministry of Education and Research (Project No. 01LG1909C).
			This research has made use of NASA's Astrophysics Data System.
		\end{acknowledgements}

	\bibliography{swsc}
\end{document}